\documentstyle[12pt,aaspp4]{article}

\def\ni{\noindent}

\def\cm{{\rm\,cm}}

\def\km{{\rm\,km}}
\def\Km{{\rm km}}

\def\g{{\rm\,g}}

\def\AU{{\rm\, AU}}
\def\Au{{\rm AU}}

\def\Hr{{\rm hr}}

\def\deg{{\rm\,deg}}
\def\Deg{{\rm deg}}

\begin{document}

\lefthead{Chiang and Brown}
\righthead{KECK KBO SURVEY}

\title{KECK PENCIL-BEAM SURVEY FOR FAINT KUIPER BELT OBJECTS}

\author{E.~I.~Chiang and M.~E.~Brown\altaffilmark{1}}
\altaffiltext{1}{Alfred P. Sloan Research Fellow}
\affil{California Institute of Technology\\
Pasadena, CA~91125, USA}

\authoremail{echiang@tapir.caltech.edu and mbrown@gps.caltech.edu}

\begin{abstract}
We present the results of a pencil-beam survey of the Kuiper Belt using the
Keck 10-m telescope.
A single 0.01 square degree field is imaged 29 times for a total integration
time of 4.8 hr.
Combining exposures in software allows the detection of Kuiper Belt Objects
(KBOs) having
visual magnitude $m_V \lesssim 27.9$. Two new KBOs are discovered.
One object having $m_V = 25.5$ lies at a probable heliocentric distance $R
\approx 33 \AU$.
The second object at $m_V = 27.2$ is located at $R \approx 44 \AU$. Both KBOs
have
diameters of about 50 km, assuming comet-like albedos of 4\%.

Data from all surveys are pooled to construct the luminosity function from $m_R
= 20$ to 27.
The cumulative number of objects per square degree, $\Sigma (< m_R)$, is fitted
to a power law
of the form $\log_{10} \Sigma = \alpha (m_R - 23.5)$, where the slope $\alpha =
0.52 \pm 0.02$.
Differences between slopes reported in the literature are due mainly to which
survey data
are incorporated in the fit, and not to the method of analysis.
The luminosity function is consistent with a power-law size distribution
for objects having diameters $s$ = 50--500 km;
$d\!N \propto s^{-q} \, ds$, where the differential size index
$q = 3.6 \pm 0.1$. The distribution is such that the smallest objects
possess most of the surface area, but the largest bodies contain the
bulk of the mass. We estimate to order-of-magnitude that $0.2 M_{\oplus}$ and
$1 \times 10^{10}$ comet progenitors lie between 30 and 50 AU.
Though our inferred size index nearly matches that derived
by \markcite{d69}Dohnanyi (1969), it is unknown whether catastrophic
collisions are responsible for shaping the size distribution.
Impact strengths may increase strongly with size from 50 to 500 km, whereas the
derivation by \markcite{d69}Dohnanyi (1969)
assumes impact strength to be independent of size. In the present-day Belt,
collisional
lifetimes of KBOs having diameters 50--500 km
exceed the age of the Solar System by at least 2 orders of magnitude,
assuming bodies consist of solid, cohesive rock.
Implications of the absence of detections of classical KBOs
beyond 50 AU are discussed.
\end{abstract}


\section{INTRODUCTION}
\label{intro}
Beyond the orbit of Neptune lies a disk of remnant planetesimals known as
the Kuiper Belt. As outlined by \markcite{jlt98}Jewitt, Luu, \& Trujillo (1998,
hereafter JLT98),
the $\sim$100 known Kuiper Belt Objects (KBOs) divide into 3 dynamical classes.

\begin{enumerate}
\item Classical KBOs reside in low eccentricity, low inclination orbits beyond
40 AU (JLT98).
They are not associated with mean motion resonances with Neptune.

\item Resonant KBOs, of which Pluto is the largest known member, have orbital
periods
commensurate with that of Neptune and are protected against close encounters
with
that planet (\markcite{m96}Malhotra 1996). They possess moderately high
eccentricities and inclinations,
possibly excited by Neptune during a transient period of orbital migration
(\markcite{jlt98}JLT98;
\markcite{m95}Malhotra 1995; \markcite{mdl99}Malhotra, Duncan, \& Levison 1999,
hereafter MDL99;
and references therein).

\item Scattered KBOs, of which 1996 TL$_{66}$ is one member
(\markcite{letal97}Luu et al. 1997),
occupy large, highly eccentric and inclined orbits,
the result of close encounters with Neptune (\markcite{dl97}Duncan \& Levison
1997).
\end{enumerate}

Of the 3 populations, the classical Kuiper Belt appears the most untouched
dynamically.
However, the record of primordial conditions preserved by the classical Belt,
as
observed today out to 50 AU, seems heavily weathered.
Recent surveys estimate the mass of the observable Kuiper
Belt within 50 AU to be a few $\times$ $0.1 M_{\oplus}$ (e.g.,
\markcite{lj98}Luu \& Jewitt 1998; this paper).
\markcite{hmw68}Hamid, Marsden, \& Whipple (1968) used the
trajectories of short-period comets to set an upper mass
limit of $1.3 M_{\oplus}$ on a smooth ring within 50 AU.
The mass of the nearby Belt
is therefore $\sim$10--100 $\times$ smaller than the $\sim$$15 M_{\oplus}$
extrapolated from the condensable material of the outer giant planets.
Mass depletion since the time of formation
is also suggested by the existence of bodies as large as Pluto.
To build bodies of this size at 36 AU within 100 Myr (the estimated
formation time of Neptune), the standard model
of pairwise planetesimal accretion requires the primordial disk
to have at least $\sim$1--10 $M_{\oplus}$ from 29 to 41 AU,
depending on the assumed sizes of seed planetesimals (\markcite{kl98}Kenyon \&
Luu 1998; cf. \markcite{sc97}Stern \& Colwell 1997).

The cause of the presumed depletion is unclear, but Neptune is considered a
prime
suspect. \markcite{dlb95}Duncan, Levison, \& Budd (1995) calculate that
Neptune, when
fixed in its present orbit, can gravitationally eject more than
90\% of the Belt mass inside 39 AU over the age of the Solar System.
About 50--90\% of the mass between 39 and 50 AU
may be depleted by gravitational perturbations alone. These simulations are
sensitive
to assumed initial eccentricities and inclinations of test particles.
Collisions are also proposed to explain the missing mass, either by
nudging objects into unstable orbits (\markcite{df97}Davis \& Farinella 1997),
or by grinding bodies down to dust to be transported by radiation
pressure (\markcite{sc97}Stern \& Colwell 1997).
In both cases, however, only the smallest KBOs may be significantly depleted.
Collisionally relaxed populations place most of their mass in the largest
bodies,
but most of their geometric cross section in the smallest members.
Collisions might therefore preferentially grind down the smallest objects,
leaving the largest bodies undisrupted and the total mass mostly intact.
This expectation is borne out in computations by \markcite{df97}Davis \&
Farinella (1997).
Important caveats for all collisional simulations of the Kuiper Belt include
oversimplified
prescriptions for the impact strengths of KBOs, reflecting our ignorance of
their internal
structure. Erosive velocities are thought to be gravitationally stirred
by Neptune within 50 AU, but physically motivated estimates of the velocity
dispersion have
yet to be made in these simulations.

However large or small, the destructive influence of Neptune on the Kuiper Belt
may be limited to within the location of its outermost 2:1 resonance at 48 AU.
This idea has led to speculation that the surface density of Belt material
rises by $\sim$2 orders of magnitude to its assumed primordial value
somewhere beyond this radius
(\markcite{s96}Stern 1996; \markcite{sc97}Stern \& Colwell 1997;
\markcite{mdl99}MDL99).
However, no classical KBO beyond 50 AU has yet been discovered. Assuming
the shape of the KBO size distribution does not change with distance, JLT98
find
by Monte Carlo simulation that their observations are consistent with
an edge to the classical Kuiper Belt at 50 AU---a ``Kuiper Cliff''.
The first theoretical constraints on the classical Belt mass beyond 50 AU
are provided by \markcite{wh98}Ward \& Hahn (1998). They find under certain
conditions
that a Belt containing $1.6 M_{\oplus}$ from 48 to 75 AU
damps Neptune's eccentricity to its current observed value of
0.009 by the action of apsidal density waves. They calculate that the addition
of 10 times more mass in this region (masses comparable to those expected in
the minimum-mass solar nebula) would reduce Neptune's eccentricity
to less than $10^{-20}$ over the age of the Solar System. In
these computations, the outer Belt is assumed to consist predominantly
of small bodies (diameters $\ll$ 140 km) so that velocity dispersions
are sufficiently low to sustain wave action.

In the absence of any direct observations of the Kuiper Belt beyond
50 AU, we undertook a pilot survey utilizing the Keck 10-m telescope.
A single 600s exposure on Keck can achieve a depth
$m_V \approx 26$, allowing objects 100 km in diameter with
comet-like albedos to be seen out to distances just beyond 50 AU.
Combining exposures in software enables the detection of such bodies inside
70 AU. Our primary aim was to constrain the KBO luminosity function
out to $m_V \approx 28$; in this goal, we succeeded.
Our principal hope was to directly image the Kuiper Belt for
the first time beyond 50 AU; this wish remains to be fulfilled at present.

Observations are described in \S\ref{observe}. Methods of data reduction
and search strategies are set forth in \S\ref{strategy}.
Results, including actual detections and our construction of the luminosity
function from $m_R = 20$ to 28, are presented in \S\ref{result}.
Implications of our results on the size, mass,
and distance distributions of KBOs are discussed in \S\ref{discuss}.
Our principal findings are summarized in \S\ref{summary}.

\section{OBSERVATIONS}
\label{observe}
Data were taken on 31 August 1997 UT using the Keck II 10-m telescope
atop Mauna Kea in Hawaii. The Low-Resolution Imaging Spectrometer
(LRIS; Oke et al. 1995) was mounted at Cassegrain focus and employed in direct
imaging mode. The plate scale on LRIS's Textronix CCD was
0.215 $\arcsec$/pixel. The camera had a useable (vignetted) field of view of
5.67 $\times$ 7.34 square arcminutes (1582 $\times$ 2048 square pixels = 0.0115
square degrees).
A standard V filter was used. The choice of V over R was motivated
by lower sky brightness and greater solar flux at V. While some KBOs have
higher reflectances at R, others also appear neutral
(Tegler \& Romanishin 1998).

We searched for KBOs in a single, relatively star-free field at opposition
[$\alpha = 22^h 54^m 54^s$, $\delta = -6\arcdeg20\arcmin34\arcsec$
(J2000)].\footnote{The field
happened to be located 43$\arcdeg$ away in ecliptic longitude from Neptune.}
Twenty-nine exposures, each 600s in
duration, were recorded of this field. Data were read out from the CCD through
two amplifiers operating simultaneously; this procedure halved the readout
time to 60s at the cost of introducing small
differences in the amount of noise between chip halves. Each frame
was offset in position by $\sim$5--100$\arcsec$ relative to other
frames; our dithering routine enabled the construction of high-fidelity
flatfields (``skyflats'') from the science data themselves
(see \S\ref{strategy1}).
Provided the Keck telescope functioned properly, our duty cycle
efficiency was nearly 90\%. Unforeseen crashes in the mirror alignment
software limited our total effective integration time to 4.8 hr over
a 6.2 hr baseline.

A Landolt field (Landolt 1992) provided photometric standards.
The seeing ranged from 0$\farcs$65 to 1$\farcs$0 full-width at half-maximum
(FWHM), with the median seeing equal~to~0$\farcs$75.

\section{DATA REDUCTION AND SEARCH STRATEGY}
\label{strategy}
Kuiper Belt Object candidates are identified by their parallax
motions (of order $\arcsec$/hr) against the fixed stars. We employed
two search methods: a simple
blinking of individual frames to visually scan for slow-moving objects,
and a deep, recombinative blinking approach which blends the search
algorithms of Gladman et al. (1998) and Cochran et al. (1995).

Observations of candidates over a single night
are insufficient to constrain orbital parameters and to prove
membership in the Kuiper Belt. Candidates might instead be
eccentric, near-Earth asteroids whose apparent motions mimic those
of true KBOs. However, as discussed by Luu \& Jewitt (1998),
the possibility of mistaken identity appears remote, since masquerading
slow-moving objects have not appeared in their many surveys to date.
We proceed on the assumption that our (small) field is likewise uncontaminated.

\subsection{Shallow Survey: Basic Blinking}
\label{strategy1}
All image processing described
in this paper was performed with the Interactive Data Language (IDL) software
package. The 29 science frames were first corrected for CCD bias and
pixel-to-pixel variations in gain (flat-fielded). For each science
frame, a tailored flatfield was constructed from the median of the
other 28 dithered science frames.\footnote{Seven additional frames
from other observations during the same night were included in the median
flatfield.}
The fact that the images were dithered ensured that each CCD pixel
sampled the flat sky several times.

Each flattened frame had its mean sky value subtracted and its flux
normalized by scaling eleven bright, unsaturated stars distributed
across the entire frame. Position offsets required to align
the dithered images were obtained by minimizing stellar residuals of frames
subtracted pairwise.

Aligned images were blinked and visually scanned for slow-moving objects.
Three images of comparable seeing, spaced about 1 hr apart, were
blinked per session. Four triplets were blinked in all, including the
first and last frames of the night.
Results of this comparatively
shallow survey are presented in \S\ref{shalsurv}.

\subsection{Deep Survey: Forward-Reverse Recombinative Blinking}
\label{strategy2}
The basic idea underlying our deep survey is simple. Images are
stacked and shifted on top of each other according to a hypothetical
KBO proper motion. The shifted stack of images is co-added to form
a recombination image. While stationary objects appear smeared
in the recombination image, an object whose motion matches that assumed
has its signal strengthened and appears as a single seeing disk.
Thus, all collected photons are used to identify KBOs too faint to
rise above the noise of an individual image.

To reduce confusion and noise in the recombination image, it is desirable to
remove non-KBO sources of emission from individual frames before co-adding.
Towards this end, we subtracted from each individual frame the median
of the other 28 (aligned) frames.\footnote{Image processing for the
deep survey began with the sky-subtracted, flux-normalized images from
the shallow survey.} Extended, stationary, low surface brightness emission
(from resolved galaxies, for example) was thereby mostly removed from
individual images. Some pixel positions did not have the full overlap of
all 29 frames because of our dithering routine; these were
purged to ensure uniform statistics.

Cosmic rays and asteroid streaks remained in the median-subtracted frames.
Substantial residuals from stationary point sources were also left
behind, a consequence of frame-to-frame seeing variations.
All three non-KBO sources of emission
were largely eliminated by clipping high-valued pixels from the shifted
stack of images. After experimenting with various schemes,
we decided to clip the 5 highest values from each column of 29 pixels
and average the remaining 24 values.\footnote{Taking the median of all 29
frames as an example of an alternative scheme generated a still noisier
background than averaging 24 frames ($\sigma_{median-29}/\sigma_{average-24}
\approx \sqrt{\frac{\pi}{2}\frac{24}{29}} \approx 1.14$).}
Columns not having the full overlap of all 29 frames due to the
shifting process were purged altogether.
Finally, to the clipped mean image we added a
positive constant frame to restore the average background level to zero.
The resulting (rectangular) array constituted our recombination image,
which appeared satisfyingly clean aside from a few well-localized
and easily recognizable residuals from bloomed stars.

The proper motion vector of a KBO is described by its
amplitude, $\mu$, and its apparent inclination angle, $\theta$, relative
to the ecliptic as seen on the CCD. Following Gladman et al. (1998),
we visually searched for KBO candidates by blinking, in any one session,
4 recombination
images corresponding to 4 successively higher amplitudes along one inclination.
Objects characteristically came into focus and then smeared as their
actual rates of motion were approached and passed.
Recombination amplitudes ranged from $\mu = 1.1$ to 6.3 $\arcsec$/hr in
steps of $\Delta \mu = 0.4 \, \arcsec$/hr.
Inclinations ranged from $\theta = -5$ to 5$\arcdeg$ in steps of
$\Delta \theta = 5\arcdeg$.\footnote{Negative (positive) $\theta$ implies
motion near a descending (ascending) node.}
These ranges cover proper motions (as seen at opposition)
of KBOs moving on prograde, circular, heliocentric orbits with
semi-major axes $R$ of 20--120 AU and actual inclinations $i$ of up to
$30\arcdeg$.\footnote{A small correction term due to the fact that
our field was $4\fdg 5$ away from opposition in ecliptic longitude was included
in
calculating these ranges.}

Roughly 130 artificial KBOs were implanted at random locations
and searched for simultaneously with true KBO candidates. Their
magnitudes were spread uniformly between $m_V = 26$ and 29,
and their orbital parameters were chosen randomly within the ranges
cited above. Their presence in recombination images trained the eye to
recognize {\it bona fide} KBOs, and their rate of recovery provided
an estimate of true KBO detection efficiency as a function of magnitude.
Differences between artificial objects' given and recovered properties
($\delta m_V$, $\delta \mu$, $\delta \theta$) provided estimates
of systematic errors in the parameters of true candidates.
Recombination spacings were just small enough to
detect artificially implanted KBOs in at least 2 recombination frames.

A list of KBO candidates was made containing
objects (including artificial ones) which
(1) focussed and de-focussed in the correct manner,
(2) appeared in at least 2 adjacent recombination images,
(3) did not appear as a single hot pixel in any 1 image,
and (4) were not situated too close to the noisy environs of
stellar/asteroidal residuals. Objects in this list had their
magnitudes and proper motions subsequently refined on
a grid of resolution $\Delta \mu = 0.1 \,\arcsec$/hr
and $\Delta \theta = 1\fdg 25$. This process involved
extracting square subframes 20 pixels wide surrounding
each candidate and recombining them on the finer grid.
Simply selecting the grid point ($\mu$, $\theta$) for
which counts inside a circular sampling aperture
were maximized proved too simplistic a procedure.
Often the maximum-count image simply pushed hot noise
pixels into our sampling aperture. In practice, we
selected the best recombination image based on visual appearance,
a well-behaved flux profile, and in the few cases where
we could not decide, maximal counts.

Without a second night to confirm the reality of our candidates,
visual surveys of this kind are more prone to false detections.
Even apart from human bias, noisy pixels may still conspire to
masquerade as slow-moving objects. To estimate the number of false
detections in our candidate list, we repeated our entire deep
search on images recombined in the reverse
direction. Reverse in this case actually means in the apparent
prograde direction, since proper motions of KBOs at opposition are
dominated by the Earth's parallax motion and must appear retrograde.
Artificial, apparently prograde objects were also randomly inserted
in individual frames and searched for in images recombined in the reverse
direction.
In this reverse survey, which suffered the same kinds of errors
as afflicted the forward survey, objects that
fulfilled the four requirements listed above and that turned out not
to be artificially inserted were deemed chance alignments
of noise. We refer to these as ``reverse survey noise
objects.''
Statistical confidence in the detection of real KBOs
demands that the number of candidates detected
in the forward survey exceed the number detected in the
reverse survey plus the uncertainty in the latter number.
In practice, we blinked recombination frames without knowing
whether they were recombined in the forward or reverse directions,
thereby avoiding another potential source of human bias.

\section{RESULTS}
\label{result}

\subsection{Shallow Survey Results}
\label{shalsurv}

\placefigure{obj1pict}
\begin{figure}
\vspace{-2in}
\plotone{f1.port.epsi}
\vspace{-0.15in}
\caption{Individual exposures of OBJ1, with time and angular scale indicated.
The object
appears in a total of 21 frames. \label{obj1pict}}
\end{figure}

One KBO was discovered by blinking individual images. The object,
hereafter OBJ1, appears at the $\sim$5.5$\sigma$ level
in 21 out of 29 frames. In the other 8 frames, light from OBJ1
had fallen off the CCD chip as a consequence of our dithering routine.
Figure \ref{obj1pict} displays our newly discovered object in 3 consecutive
exposures,
and Table \ref{objprop} summarizes its measured and inferred properties.
Its motion over 6.2 hr is consistent with being uniform; a best-fit
line through centroid positions yields proper motion parameters
$\mu = 3.83 \,\arcsec$/hr and $\theta = -1\fdg 0$. The corresponding
heliocentric distance and inclination for an assumed circular
orbit are $R = 32.9$ AU and $i = 4\fdg 5$. With a
measured visual magnitude of
$m_V = 25.5 \pm 0.3$ (1$\sigma$ dispersion among 21 measurements),
the object is $56 \pm 6$ km in diameter, assuming it has
a comet-like visual albedo of 0.04 (Allen 1973).

\placetable{objprop}
\begin{deluxetable}{cccccccccc}
\tablewidth{0pc}
\tablecaption{Properties of Detected KBOs\label{objprop}}
\tablehead{
\colhead{Object} & \colhead{$m_V$\tablenotemark{a}} &
\colhead{$\mu$\tablenotemark{b} $(\arcsec /\Hr)$} &
\colhead{$\theta$\tablenotemark{b} $(\Deg)$} &  \colhead{$R$\tablenotemark{c}
$(\Au)$} &  \colhead{$i$\tablenotemark{c} $(\Deg)$} &
\colhead{$s$\tablenotemark{d} $(\Km)$} & \colhead{$\eta$\tablenotemark{e}} &
\colhead{$A$\tablenotemark{e} (deg$^2$)} &
\colhead{Method}}

\startdata
OBJ1 & 25.5 & 3.83 & -1 & 32.9 & 4.5 & 56 & 100\% & 0.0102 & Shallow \nl
OBJ2 & 27.2 & 2.92 & 0  & 43.9 &  0  & 46 & 98\%  & 0.009  & Deep \nl
\tablenotetext{a}{Measured visual magnitude, uncertain by 0.3 (0.22) mag for
OBJ1 (OBJ2).}
\tablenotetext{b}{Measured proper motion amplitude and angle relative to
ecliptic on CCD, respectively. For OBJ1 (OBJ2), uncertainties are 0.02 (0.05)
$\arcsec$/hr and 1 (2.3) degrees.}
\tablenotetext{c}{Inferred heliocentric distance and inclination, respectively,
for an assumed circular orbit.}
\tablenotetext{d}{Inferred diameter, assuming a visual albedo of 4\%.}
\tablenotetext{e}{Detection efficiency and area searched, respectively.}
\enddata
\end{deluxetable}

The area of sky covered by our shallow survey was $A_s$ = 0.0102 square
degrees,
after correcting approximately for dithering losses (-11\%) and
area taken up by bright stars and galaxies (-0.5\%).
A discussion of the cumulative luminosity function is reserved
for \S\ref{lumfunc}.

\subsection{Deep Survey Results}
\label{deepsurv}

\subsubsection{Artificial Object Recovery}
\label{artobj}

\begin{figure}
\plotone{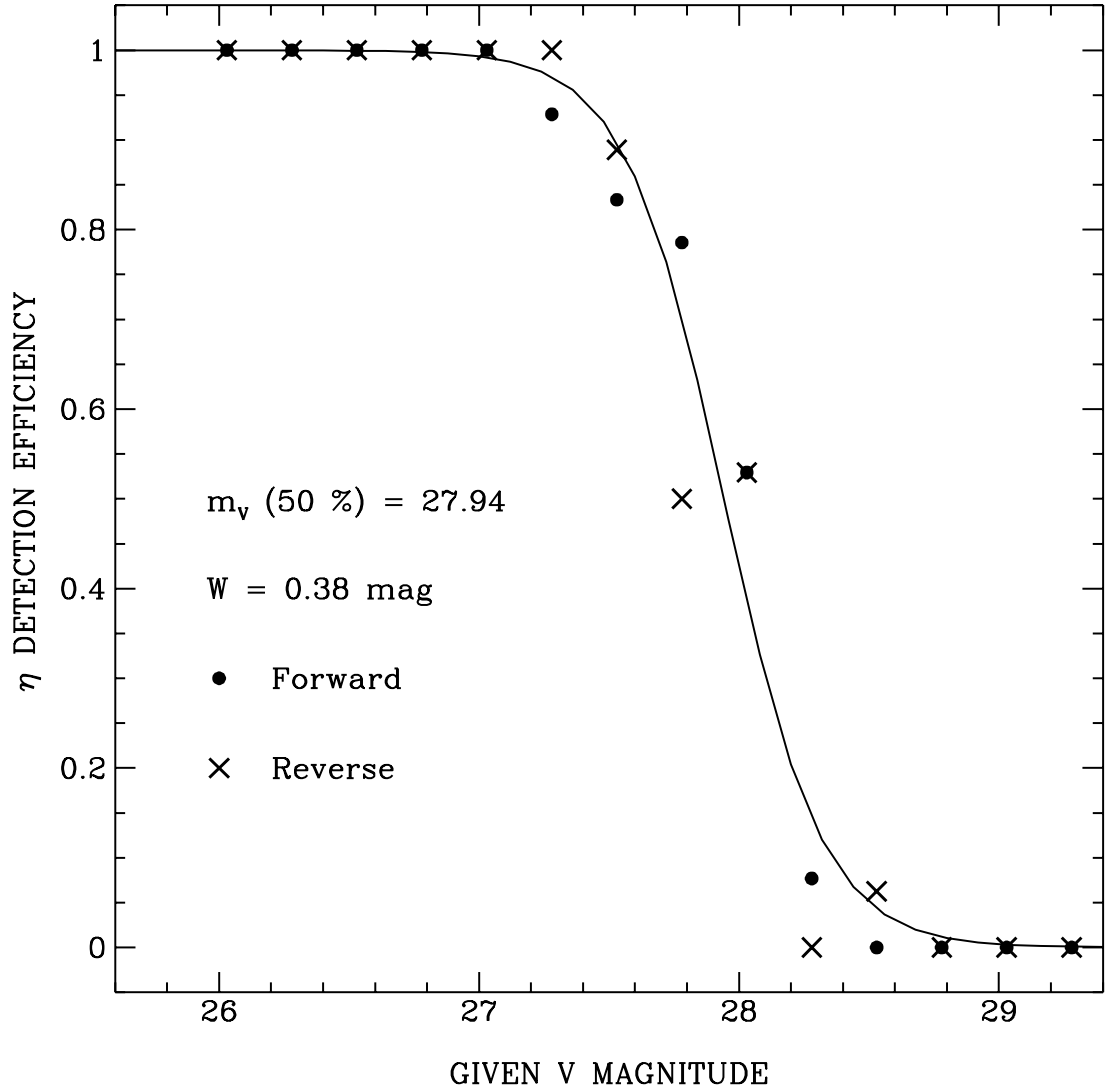}
\caption{Rate of recovery of artificially implanted objects versus their given
magnitude.
Detection efficiencies from forward and reverse surveys are averaged and fitted
to equation (\protect{\ref{deteff}}),
shown as a solid line. \label{effdet}}
\end{figure}

After refining estimates of candidates' magnitudes and proper motions,
we culled artificial objects from the candidate list.
Figure \ref{effdet} displays our recovery rate $\eta$ of artificial objects
as a function of their given $m_V$, for both forward and reverse
surveys. In both surveys, the rate of recovery was similar, falling
from 100\% near $m_V = 27.3$ to 0\% at $m_V = 28.4$. The datasets
were combined and fitted to the function

\begin{equation}
\eta (m_V) = \frac{1}{2} \left[ 1 - \tanh \left( \frac{m_V - m_V{\rm(50\%)}}{W}
\right) \right]
\label{deteff}
\end{equation}

\ni
(Gladman et al. 1998). The fit yields a detection efficiency which
falls to 50\% at $m_V$(50\%) = 27.94, over a characteristic width
$W = 0.38$ mag. Satisfyingly, $m_V$(50\%) is only 0.04 mag brighter than
the nominal 3$\sigma$ limit obtained by reducing the noise
of an individual image by $\sqrt{24}$.

\placefigure{effdet}

\begin{figure}
\plotone{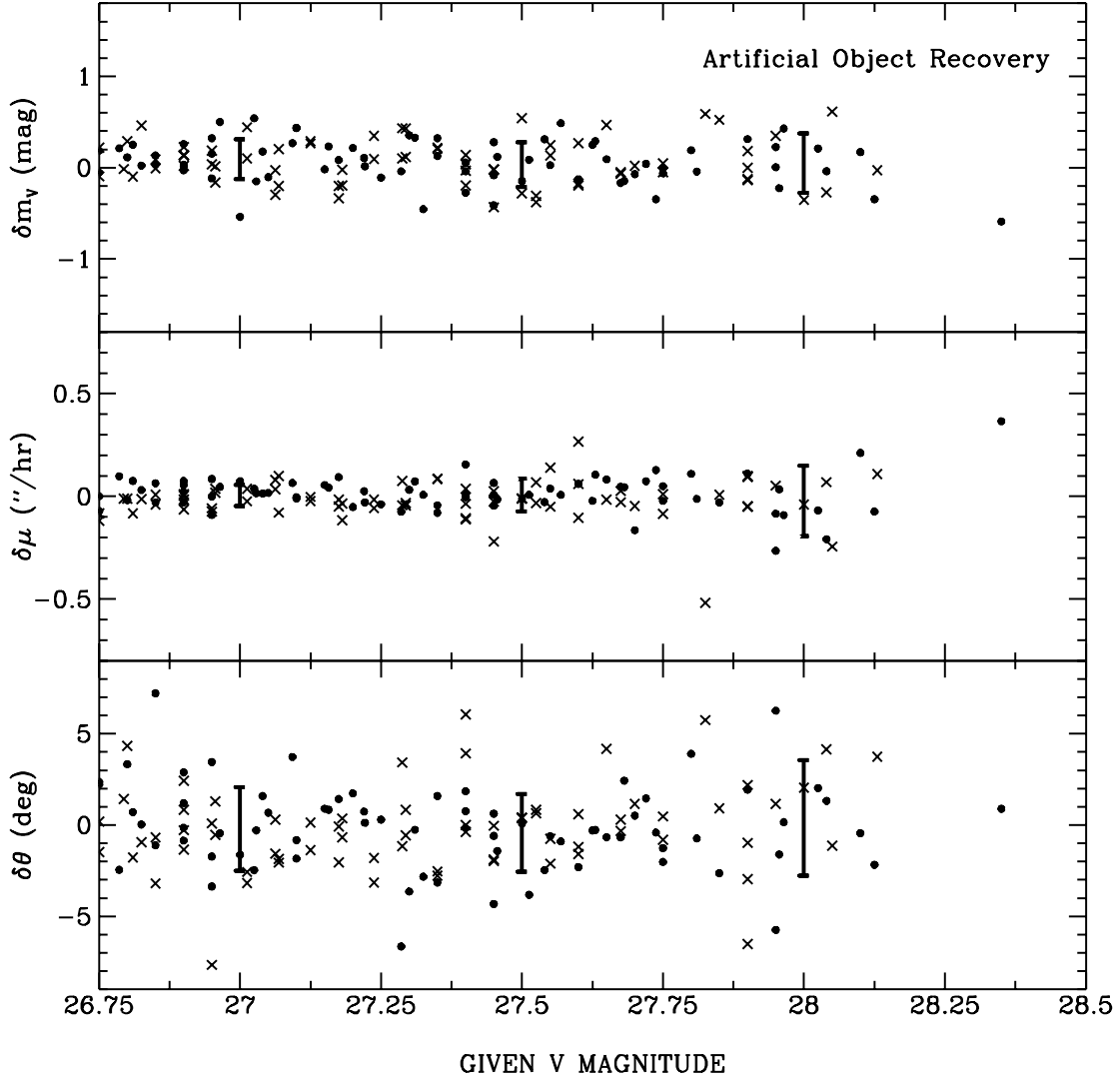}
\caption{Differences between given and recovered properties of artificial
objects. Solid circles
indicate objects recovered in the forward survey, and crosses denote objects
recovered in the reverse
survey. Error bars reflect $\pm$1$\sigma$ dispersions in bins of width 0.5 mag.
Since average differences
are consistent with being zero, we conclude that our measurements of $m_V$,
$\mu$, and $\theta$ are not
biassed. We adopt the dispersions to be our measurement uncertainties.
\label{refinerr}}
\end{figure}

Figure \ref{refinerr} plots $\delta m_V$, $\delta \mu$, and $\delta
\theta$---differences
between artificial objects' given and recovered properties---versus their
given magnitude. To clarify possible trends with increasing magnitude,
we also plot averages and standard deviations within bins of width
0.5 mag; these points are positioned at the centers of each bin.

\placefigure{refinerr}

We note first that there is no significant bias in our estimation
of parameters; averages $\overline{\delta m_V}$, $\overline{\delta \mu}$,
and $\overline{\delta \theta}$ are consistent with being zero.
The scatter, however, is significant. We adopt the scatter in
$\delta m_V$ as our estimate of the uncertainty in true candidates' magnitudes;
the 1$\sigma$ dispersion increases from 0.22 mag near $m_V = 27$
to 0.33 mag near $m_V = 28$. The analogous 1$\sigma$ uncertainty
in $\mu$ ranges from 0.05 to 0.17 $\arcsec$/hr, and
the 1$\sigma$ uncertainty in $\theta$ ranges from $2\fdg 3$ to $3\fdg 2$.
These results for $\mu$ and $\theta$ appear reasonable.
A difference of $\Delta \mu = 0.1$ $\arcsec$/hr
over a time $\Delta t = 6.2$~hr smears images
by $\Delta \mu \times \Delta t = 0\farcs 6$---about half the
value of the worst seeing during our observations.
A difference of $\Delta \theta = 2\fdg 5$ at a fixed, typical amplitude of
$\mu = 3$ $\arcsec$/hr smears images over
a comparable distance: $\mu \times \Delta \theta \times \Delta t = 0\farcs 8$.

\subsubsection{True Object Discovery and Upper Limits}
\label{truecan}
\begin{figure}
\plotone{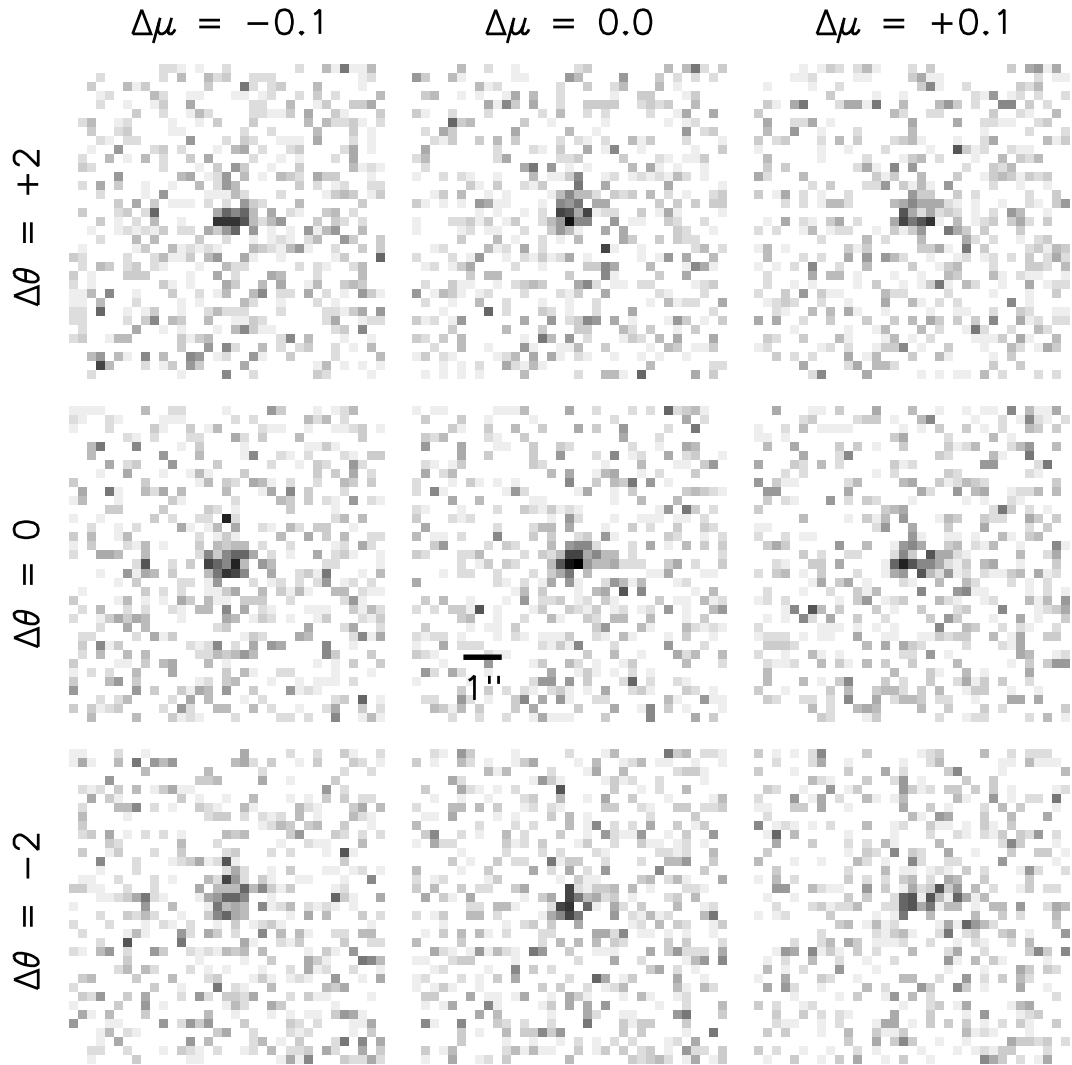}
\caption{Recombination images of OBJ2. The central image is the
best recombination image. Surrounding it are images recombined at
adjacent points on the ($\mu$, $\theta$) grid. Each panel to the right advances
$\Delta \mu = 0.1 \arcsec/\Hr$.
Each panel towards the top of the page advances $\Delta \theta = 2 \arcdeg$.
\label{obj2pict}}
\end{figure}

A second KBO, hereafter OBJ2, was discovered by blinking recombination
frames. Figure \ref{obj2pict} presents the best recombination image of OBJ2,
surrounded
by images of the same object recombined at adjacent points on the
($\mu$, $\theta$) grid. Its smearing pattern is identical to those of
artificially planted objects having similar motions. Properties of OBJ2
are summarized in Table \ref{objprop}. Its visual magnitude is $m_V = 27.22 \pm
0.22$,
and its proper motion parameters are $\mu = 2.92 \pm 0.05$ $\arcsec$/hr
and $\theta = 0\arcdeg \pm 2\fdg 3$. Alternative recombination frames
for OBJ2 were constructed by clipping the top 1 pixel out of each column
of 29 pixels and then averaging the remaining 28 values. Exactly the same
parameters for OBJ2 were obtained. If we assume a comet-like visual
albedo of $p_V = 0.04$, these measurements are consistent with those of an
object $46 \pm 6$ km
in diameter, occupying a circular, uninclined orbit of semi-major
axis $R = 43.9 \pm 0.8$ AU.

\placefigure{obj2pict}

\begin{figure}
\plotone{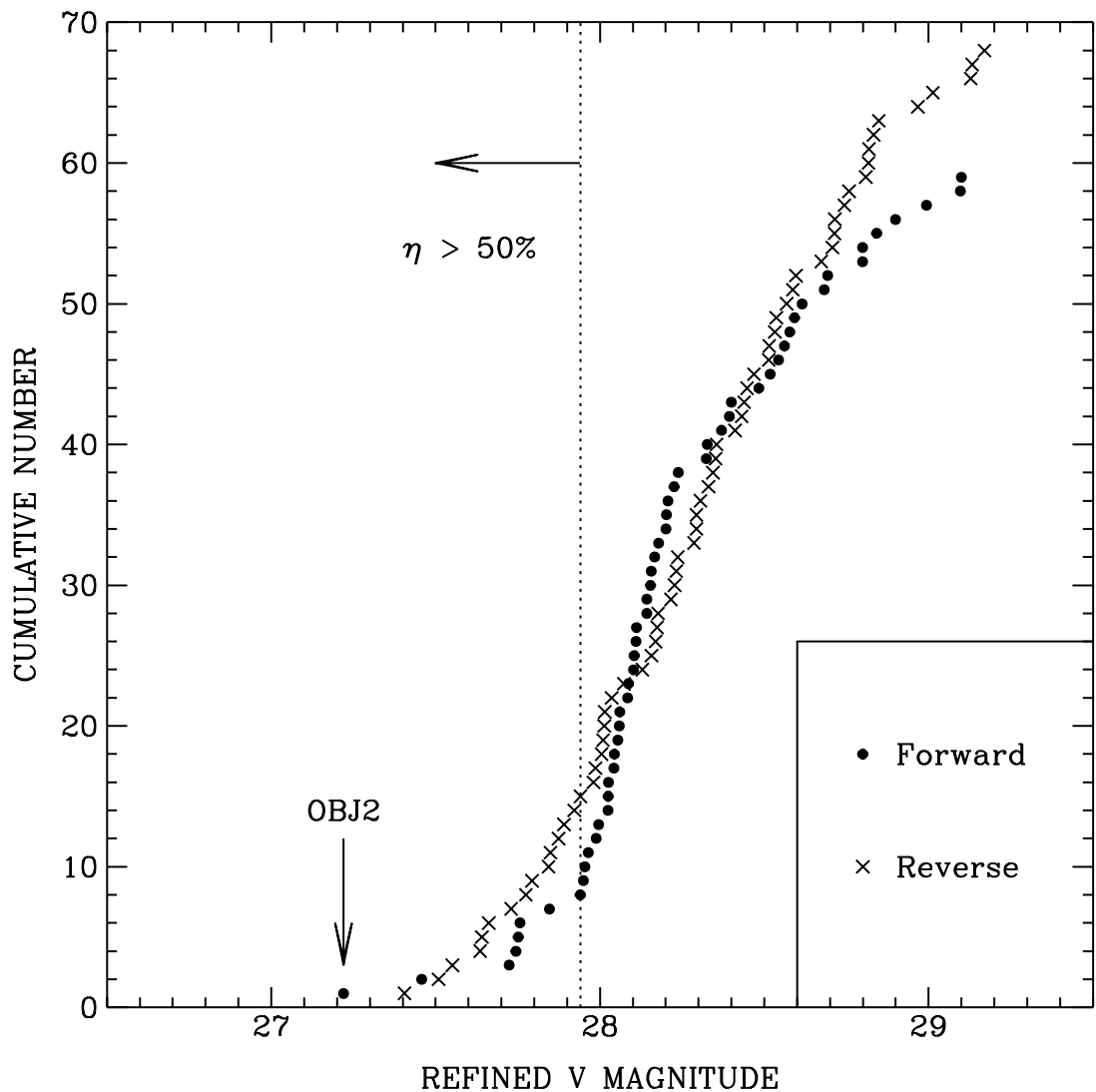}
\caption{Comparison of the number of KBO candidates found in the deep forward
survey and the number of noise objects detected in the reverse survey.
At $m_V < 27.4$, the only object detected is OBJ2, and no noise object is
bright enough to confuse the identity of OBJ2 as a true KBO. At $m_V > 27.4$,
the number of KBO candidates never significantly exceeds the number of
false alarms, and we can only compute upper limits on the sky density.
\label{forrev}}
\end{figure}

Confidence in the reality of OBJ2 is further bolstered by Figure \ref{forrev},
in which we compare cumulative numbers of objects detected in
forward and reverse surveys. No false alarm went off in the reverse
survey at the magnitude of OBJ2; the object
distinguishes itself as the brightest detection at 5.5$\sigma$.

\placefigure{forrev}

By contrast, we view all candidates in the forward survey fainter
than $m_V = 27.4$ as false detections, partly because their numbers do not
exceed those in the reverse survey. No object in both
surveys is as visually convincing as OBJ2; many other candidates
vanished at several (but not all) adjacent
recombination gridpoints. Moreover,
regarding the last of the 4 search criteria set forth in \S\ref{strategy2},
it was occasionally unclear when an object was ``too close'' to
a smeared stellar residual. Thus, some of our detections fainter than $m_V =
27.4$
undoubtedly arise from the confusing noise of bloomed stars (OBJ2 is far
removed from any such noise). We use the population
of noise objects detected in the reverse survey to set upper limits
on the cumulative sky density of KBOs fainter than $m_V = 27.4$. Details of
this
calculation follow in the next section.

The area searched in our deep survey is less than that of our shallow
survey because of the shifting process. Areal losses ranged from 7--13\%
depending on the value of $\mu$. To simplify the analysis,
we adopt an average loss of -10\%; the error introduced is
negligible compared to Poisson uncertainties in the number
of objects detected. Corrected for additional losses
due to stellar/asteroidal residuals (-1\%),
our deep survey area equals 0.009 square degrees $\equiv A_d$.

\subsection{Cumulative Luminosity Function}
\label{lumfunc}
\begin{figure}
\plotone{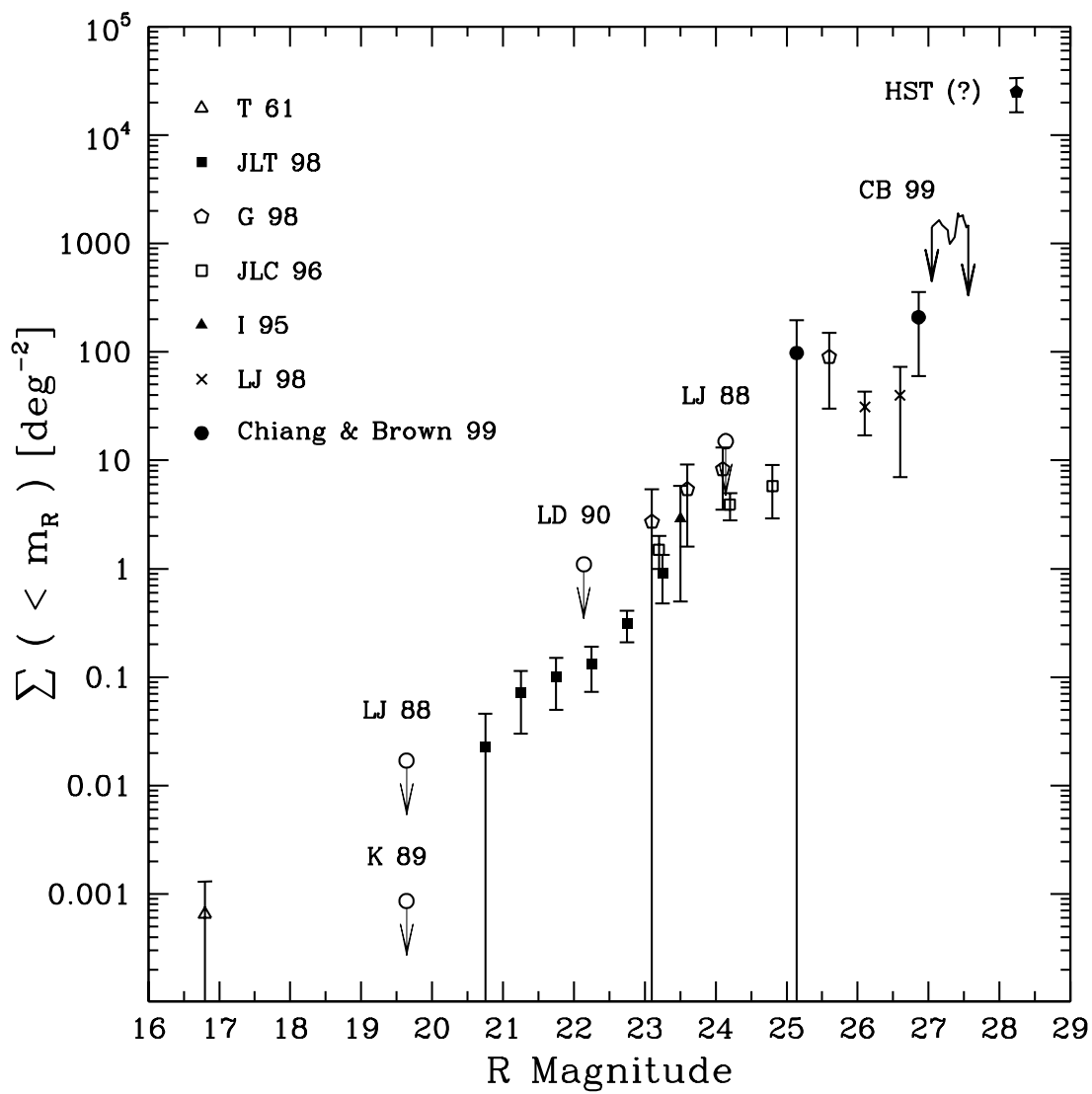}
\caption{Independent estimates of the cumulative sky density of KBOs as made by
various groups. Abbreviations for surveys are defined in the text and in the
references.
Upper limits from this paper (CB99) are computed at the 99.99\% (``4$\sigma$'')
confidence
level. Upper limits from other surveys
are published values at the 99\% confidence level. \label{clfevery}}
\end{figure}

Figure \ref{clfevery} displays our estimates of the cumulative KBO
sky density, $\Sigma (< m_R)$, together with estimates made
by various other groups. We emphasize that each survey's
points represent estimates made independently of all other
groups; i.e., survey areas have not been added.\footnote{The
one exception to independence involves the
points from \markcite{lj98}Luu \& Jewitt (1998), which presumably
incorporate data from their previous surveys.} Surveys conducted
in V were included by assuming a solar color, V-R = 0.36,
corresponding to a neutrally reflective KBO (red albedo $p_R = p_V$).

\placefigure{clfevery}

{}From our detection of OBJ1, we independently estimate
$\Sigma (m_R < 25.14) = 98 \pm 98 (1\sigma)$ objects/deg$^2$. Combining this
result with our detection of OBJ2, which is the faintest KBO detected to date,
we estimate $\Sigma (m_R < 26.86) = 209 \pm 149 (1\sigma)$ objects/deg$^2$,
where Poisson
uncertainties have been added in quadrature.

Upper limits are derived at fainter magnitudes as follows.
We assume that in the forward survey, the occurrence of noise objects
plus real KBOs is Poissonian. The expected
mean number of forward survey candidates brighter than magnitude $m$ in survey
area $A_d$ equals $N_{\rm{Noise}}(<m) + \langle \eta A_d \rangle \Sigma(<m)$,
where $N_{\rm{Noise}}(<m)$ is the mean cumulative number of noise objects,
and $\langle \eta A_d \rangle$ is the efficiency-weighted survey area.
We take $N_{\rm{Noise}}(<m) = N_R(<m)$, where $N_R(<m)$
is the cumulative number of reverse survey noise objects found.
Given the number of forward survey candidates
that we actually detected, $N_F (<m)$, we ask what minimum value of $\Sigma
(<m)$
can be ruled out at the 99.99\% confidence level ($\sim$``4$\sigma$'' in
Gaussian parlance):

\begin{equation}
\frac{(N_R + \langle \eta A_d \rangle \Sigma)^{N_F} \exp -(N_R + \langle \eta
A_d \rangle \Sigma)}{N_F !} = 10^{-4} \, ,
\label{ulim}
\end{equation}

\noindent an implicit equation for $\Sigma$ where the magnitude dependence has
been dropped
for compactness. For reverse survey noise objects brighter than $m_V(50\%)$, we
take $\langle \eta A_d \rangle = 0.7 \times A_d$.
We do not calculate upper limits for $m_V > m_V(50\%)$, since
our detection efficiency falls rapidly to zero past that magnitude (see Figure
\ref{effdet}).
Upper limits on $\Sigma$ computed using equation (\ref{ulim}) are plotted in
Figures
\ref{clfevery} and \ref{clfuni}.

While Figure \ref{clfevery} summarizes the history of KBO surveys, quantitative
results such as the slope of the luminosity function (or even the degree
to which $\Sigma$ resembles a single-slope power law) are better
extracted from a fairer pooling of the data. To this end,
we imagine the areas from all surveys as being combined into one giant frame
over which
the detection efficiency varies. At magnitude $m_i$ of a detected KBO,

\begin{equation}
\Sigma (<m_i) = {\displaystyle \sum_{j=1}^i} \,\,\,\,\frac{1}{{\displaystyle
\sum_{k=1}^n} \eta_k(m_j) A_k} \, ,
\label{pool}
\end{equation}

\ni where $m_j$ is the magnitude of the $j^{th}$ brightest KBO,
$\eta_k \times A_k$ is the efficiency-weighted area of the $k^{th}$ survey,
and $n$ is the total number of surveys. Most surveys have published
efficiency functions. Exceptions include the Mauna Kea-Cerro Tololo survey
of \markcite{jlc96}Jewitt, Luu, \& Chen (1996, hereafter JLC96), the
Keck survey by \markcite{lj98}Luu \& Jewitt (1998, hereafter LJ98), the
McGraw-Hill
CCD survey by \markcite{lj88}Luu \& Jewitt (1988, hereafter LJ88), and
the U.S. Naval Observatory survey by \markcite{ld90}Levison \& Duncan (1990,
hereafter LD90).
For data from JLC96, we assume
$\eta$ behaves in a similar manner to that described in their
companion paper I \markcite{jl95}(Jewitt \& Luu 1995, hereafter JL95);
i.e., $\eta$ is assumed to fall linearly from 100\% to 0\%
over 0.7 mag centered on published values of $m_R$(50\%).
For data from LJ98, we obtained $\eta$
by consulting the lead author \markcite{l99}(Luu 1999).
For the 2 remaining older surveys [which detected no KBOs, but
which nonetheless contribute slightly to the total survey area in equation
(\ref{pool})],
we adopted Heaviside step functions centered at $m_R = 24$ (LJ88) and $m_R =
22.14$ (LD90).
We have verified that the conclusions of our paper are not affected by
how we incorporate the latter 2 surveys. No photographic survey was included in
the pool.

\begin{figure}
\plotone{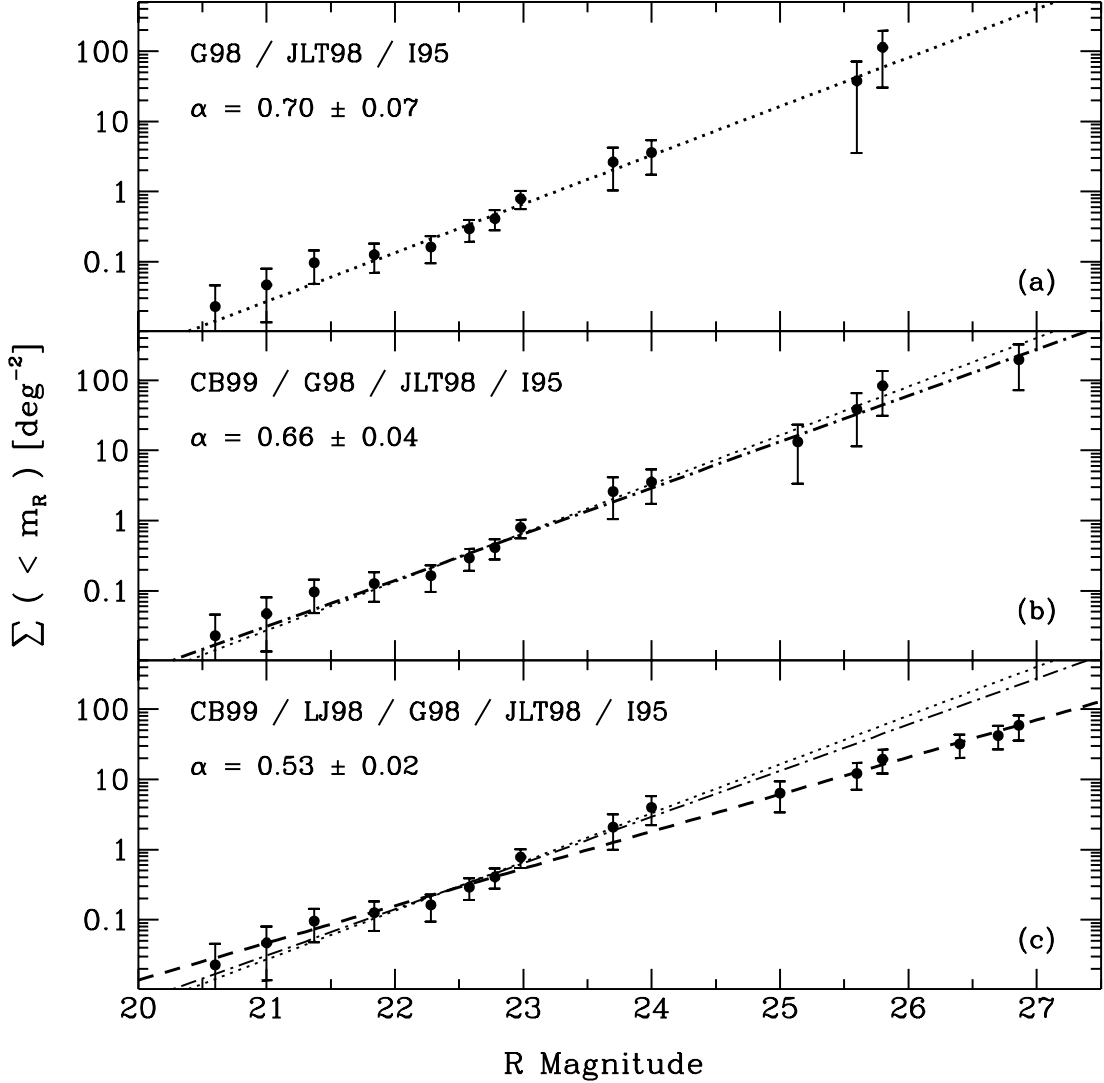}
\caption{Cumulative sky density obtained by pooling surveys
according to equation (\protect{\ref{pool}}). From panels (a) through (c),
successively more surveys are pooled, as indicated by the growing list
of acronyms at the top of each panel. Data are fitted by least-squares,
with fits from preceding panels plotted for comparison. In panel (a),
the fitted slope $\alpha$ of the luminosity function is identical to that
derived
using a maximum likelihood analysis by \protect{\markcite{g98}}Gladman et al.
(1998). In panel (c),
the slope $\alpha$ decreases significantly when data from LJ98 are included.
\label{clftri}}
\end{figure}

\begin{figure}
\plotone{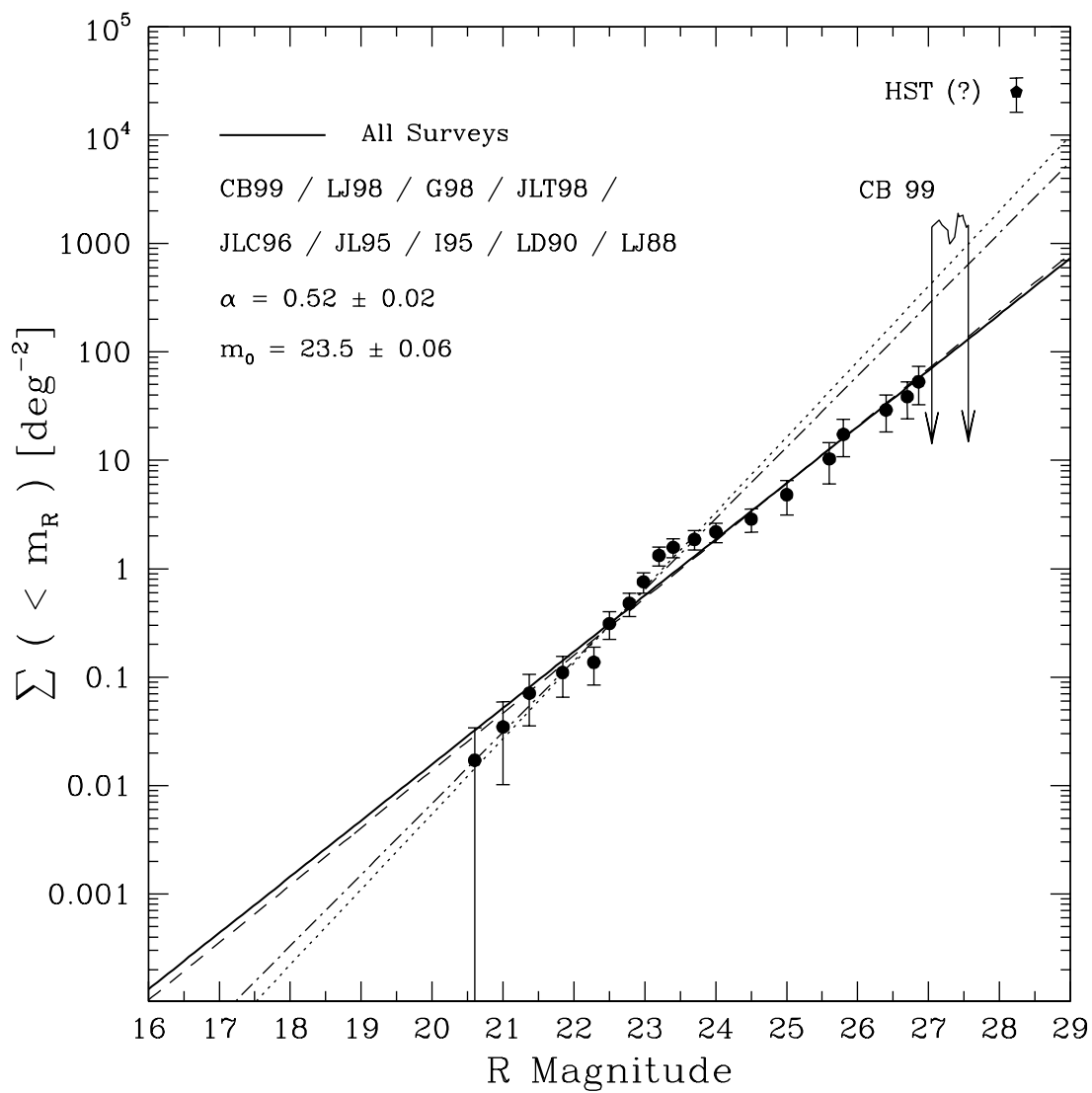}
\caption{Cumulative sky density obtained by pooling all surveys.
The solid line is the power law fitted to all survey data. Other lines are fits
from Figure \protect{\ref{clftri}}, re-plotted here for comparison. Neither the
HST datum nor \protect{\markcite{t61}}Tombaugh's (1961) datum is included in
any fit.
Data from JL95 and JLC96 reinforce the shallow slope forced by data from LJ98.
\label{clfuni}}
\end{figure}

\placefigure{clftri}

Figure \ref{clftri} displays the results of pooling datasets according to
equation
(\ref{pool}). To clarify the roles played by individual surveys,
we pool an incrementally larger number of surveys in Figures \ref{clftri}a
through
\ref{clftri}c. In these and subsequent plots, magnitudes of individual points
are identical to magnitudes $m_i$ of individual KBOs. However,
only points separated by at least $\sim$0.2 mag are plotted; this represents
a minor smoothing of the dataset, but is still preferable to imposing
arbitrary bin boundaries. Error bars reflect Poisson counting statistics.

Consider first Figure \ref{clftri}a, which incorporates data from
\markcite{izt95}Irwin, Tremaine, \& Zytkow (1995),
\markcite{jlt98}Jewitt, Luu, \& Trujillo (1998),
and \markcite{getal98}Gladman et al. (1998) (hereafter I95,
JLT98, and G98, respectively). These constitute the 3 surveys
preferred by \markcite{getal98}Gladman et al. (1998),
excluding upper limit data. The points are well described by a power
law, written in conventional notation as

\begin{equation}
\Sigma (<m_R) = 10^{{\displaystyle \alpha (m_R - m_0)}} \, ,
\label{powerlaw}
\end{equation}

\ni where slope $\alpha$ and reference magnitude $m_0$ are fitted
parameters. A least-squares fit to these 3 surveys alone yields $\alpha = 0.70
\pm 0.07$,
$m_0 = 23.3 \pm 0.1$. These values coincide with those derived
using a maximum likelihood analysis by \markcite{getal98}Gladman et al. (1998);
see their Figure 6c. We realize that
least squares is not the preferred statistic for data whose
errors are not Gaussian and which are correlated from point to point.
However, the agreement between our result and G98's suggests that differences
between slopes derived by various groups are due mainly to which
surveys are kept and which are neglected, and not to the method of analysis.
This will be borne out in what follows.

In Figure \ref{clftri}b, we incorporate our survey (hereafter CB99) into the
pool.
The slope is lowered slightly to $\alpha = 0.66 \pm 0.04$,
but the change is negligible over the range of observed magnitudes.
On the basis of these 4 surveys alone, our data extend the
$\alpha \approx 0.7$ law to $m_R \approx 26.9$.

In Figure \ref{clftri}c, we fold in the Keck survey of \markcite{lj98}LJ98.
The observed faint end of the luminosity
function is suppressed by the weight of their relatively sparsely
populated fields. The luminosity function still resembles a single-slope
power law, but the refitted slope is substantially
shallower; $\alpha = 0.53 \pm 0.02$,
coincident with the value given by LJ98.
Though the LJ98 fields have a few times fewer objects at
$m_R \approx 26.5$ compared to our CB99 fields, discrepancies are at the
$\sim$1$\sigma$ level or less; uncertainties in our points (see Figure
\ref{clftri}b)
are large because we detect only 2 objects.

\placefigure{clfuni}

Finally, in Figure \ref{clfuni}, the remaining surveys by JL95, JLC96, LJ88,
and LD90
are assimilated into the pool. The fitted luminosity function
hardly changes; for this final
pool, $\alpha = 0.52 \pm 0.02$ and $m_0 = 23.5 \pm 0.06$.
We note that shallow values for the fitted slope
depend not only on surveys by JL95 and JLC96, but also on recent data
from LJ98 (see Table 2).
Some crude, model-dependent considerations of why
values of $\alpha < 0.6$ might be preferred are given in \S\ref{size}.

The claimed Hubble Space Telescope (HST) detection
of $2.5 \times 10^4$ objects/deg$^2$ at $m_R \approx 28.2$
\markcite{clsd95}(Cochran et al. 1995) lies $\sim$10 times
above the steepest extrapolation, and $\sim$100 times above
the extrapolation derived from all surveys combined.
\markcite{bkl97}Brown, Kulkarni, \& Liggett (1997) independently
suggest on statistical grounds that the detections are erroneous.
The increasing difficulty of reconciling the ground-based observations
with the HST claim appears to support this suggestion.

\section{DISCUSSION}
\label{discuss}

\subsection{Size, Surface Area, and Mass Distributions}
\label{size}

We consider a power-law differential size
distribution with index $q$, $d\!N(s) \propto s^{-q} \, ds$,
where $d\!N(s)$ is the number density of objects having diameters between $s$
and $s + ds$. If all observed KBOs had the same albedo and were at the
same heliocentric distance, the measured slope $\alpha$ of the
cumulative luminosity function would imply a unique size index, viz.

\begin{equation}
q = 5.02 \alpha + 1 \, .
\label{simple}
\end{equation}

\ni
This relation is straightforward to derive and is first given by
\markcite{itz95}I95. The assumption of uniform distance is not a
bad one, since KBOs detected to date have present-day heliocentric
distances between 30 and 50 AU; adopting a geometric mean distance for all
objects mis-estimates sizes by at most a factor of 5/3. This is
less than the possible factor of 4 uncertainty in size introduced by
the unknown albedo, which might range from 4\% (Comet Halley) to
60\% (Pluto). \markcite{jlt98}JLT98 employ Monte Carlo models which
incorporate more realistic distance distributions to extract
the size index from the measured luminosity function. Their
best-fit $q = 4.0 \pm 0.5$ agrees with the value obtained by
inserting their measured $\alpha = 0.58 \pm 0.05$
into equation (\ref{simple}). We shall use equation (\ref{simple})
to calculate $q$ from $\alpha$ below, keeping in mind that
such $q$'s may be uncertain by $\pm$0.5.

\placetable{alphaq}
\begin{deluxetable}{clcl}
\tablewidth{0pc}
\tablecaption{Measured $\alpha$\tablenotemark{a} $\,$ and Inferred
$q$\tablenotemark{b} \label{alphaq}}
\tablehead{
\colhead{$\alpha$} & \colhead{Source} & \colhead{$q$} &
\colhead{Implication}}

\startdata
0.52 & All data (Fig. 8) & 3.6 & Mass in Largest Bodies \nl
0.53 & Omit JL95, JLC96 (Fig. 7c) & 3.7 & Mass in Largest Bodies \nl
0.57 & Omit LJ98 & 3.9 & Mass in Largest Bodies \nl
0.66 & Omit JL95, JLC96, LJ98 (Fig. 7b) & 4.3 & Mass in Smallest Bodies \nl
0.70 & Omit JL95, JLC96, LJ98, CB99 (Fig. 7a) & 4.5 & Mass in Smallest Bodies
\nl
\tablenotetext{a}{Power-law slope of cumulative luminosity function; see
equation (\ref{powerlaw}).}
\tablenotetext{b}{Differential size index derived from $q = 5.02 \,\alpha + 1$,
which assumes uniform albedo and distance. See \S\ref{size} for discussion.}
\enddata
\end{deluxetable}

Table \ref{alphaq} summarizes possible values of the sky density slope $\alpha$
and the
size index $q$ and their implications.
Depending on which surveys are incorporated, $q$ takes values from
3.6 to 4.5. We compare these values to those of erosive
disks in our Solar System. Main-belt asteroids are inferred to obey
$q \approx 3.3$ in the diameter range 3--30 km
(\markcite{dgj98}Durda, Greenberg, \& Jedicke 1998).
A value of $q = 3.5$ corresponds to a quasi-steady-state
population for which catastrophic collisions
grind as much mass per time into every size bin as they grind out,
as first derived by \markcite{d69}Dohnanyi (1969).
The derivation further assumes that critical
specific energies for shattering and dispersal are independent of
size.\footnote{The critical
specific energy for shattering, $Q^{*}_S$, is defined as the energy per unit
target mass required
to produce a fragment with 50\% the mass of the original target. It is smaller
than
$Q^{*}_D$, the energy per unit target mass required to disperse such
fragments to infinity \markcite{mr97}(Melosh \& Ryan 1997).} This
is a fair assumption for solid rocky targets smaller than $\sim$10 km in
diameter for which
internal compression due to self-gravity is negligible.
For asteroids greater than 30 km in diameter, there are significant deviations
from the
$q \approx 3.5$ law, with slopes ranging from $q \approx 2$ to 4.5
(\markcite{dd97}Durda \& Dermott 1997).
These deviations likely result from variations of the impact
strength with size, as caused by self-gravitational effects
(\markcite{dgj98,mr97}Durda, Greenberg, \& Jedicke 1998; Melosh \& Ryan 1997).
Saturn's ring particles crudely fit $q \approx 3.3$ in the size
range of a few centimeters to a few meters based on Voyager radio
occultation data \markcite{metal83}(Marouf et al. 1983), though
values between 2.8 and 4.0 cannot be completely ruled out
\markcite{cetal84,wetal84}(Cuzzi et al. 1984, Weidenschilling et al. 1984).
Ring optical depths are sufficiently high that particles
have suffered many erosive collisions over the age of the Solar
System, so that their size distribution no longer purely reflects initial
conditions \markcite{bgt84}(Borderies, Goldreich, \& Tremaine 1984).

For $q > 3$ ($\alpha > 0.4$), surface areas (geometric optical depths)
are dominated by the smallest bodies. All current estimates of $q$ imply
that this is the case for the Kuiper Belt.

For $q < 4$ ($\alpha < 0.6$), the total mass is dominated by the largest
bodies. If we combine all surveys, we infer a size index $q = 3.6 \pm 0.1$.
We use this $q$ to make an order-of-magnitude estimate of the mass in the
observable
Kuiper Belt. Nearly all KBOs in the surveys
we have considered have inferred diameters $s \gtrsim 50 \, (0.04 / p_R)^{1/2}
\, \km$.\footnote{The only
exception is KBO K3, for which $s = 23 \km$ \markcite{lj98}(Luu \& Jewitt
1998).}
At limiting magnitude $m_R = 27$ (the V-R adjusted magnitude above which false
alarms prevent additional KBO detections in our deep survey), objects having
$s \gtrsim 50 \km$ can be seen out to distances of 48 AU.
For values of $q$ and $\Sigma (m_R < 27)$
derived by combining all survey data, the total mass of the Kuiper
Belt out to 48 AU is

\begin{eqnarray}
M_{Belt} (a < 48 \AU) & \approx & 0.22 \left( \frac{\Sigma (m_R < 27)}{53
\deg^{-2}} \right) \left( \frac{A_{KB}}{10^4 \deg^2} \right) \left(
\frac{\rho}{2 \g\cm ^{-3}} \right) \times \nonumber \\
 & & \left( \frac{0.04}{p_R} \right)^{1.3} \left( \frac{s_{max}}{2000 \km}
\right)^{0.4} M_{\oplus} \, .
\label{mass}
\end{eqnarray}

\ni Here $A_{KB}$ is the solid angle subtended by the Kuiper Belt (taken to
extend $\pm 15\arcdeg$
in ecliptic latitude), $\rho$ is the internal mass density of KBOs,
and $s_{max}$ is the diameter of
the largest body in the distribution (taken to be similar to
Pluto).\footnote{Our calculation
ignores the fact that some surveys observe $\pm$90$\arcdeg$ away in ecliptic
longitude from
Neptune where Plutinos (KBOs in 3:2 resonance with Neptune) reach perihelion
(\markcite{m96}Malhotra 1996).
These surveys might be expected to find an unrepresentatively high sky density
of KBOs.
In fact these surveys (JL95, JLC96) find lower sky densities than other
surveys; see \S\ref{lumfunc}
and section 5.1 of \markcite{g98}G98.}
Our rough estimate of $\sim$0.2 $M_{\oplus}$ is consistent with
the upper limit of 1.3 $M_{\oplus}$ within 50 AU
derived by \markcite{hmw69}Hamid, Marsden, \& Whipple (1968) using measured
cometary orbits
(see \markcite{w95}Weissman 1995 and \markcite{mdl99}MDL99 for a discussion of
upper mass limits).
Note that this model predicts the existence of $\sim$10 more Pluto-sized
objects in the nearby
Kuiper Belt.

The number of 1-10 km sized comet progenitors in the Kuiper Belt may be
similarly
estimated;

\begin{equation}
N_{Comet} (a < 48 \AU) \approx 1.4 \times 10^{10} \left( \frac{\Sigma (m_R <
27)}{53 \deg^{-2}} \right) \left( \frac{A_{KB}}{10^4 \deg^2} \right) \left(
\frac{0.04}{p_R} \right)^{1.3} \left( \frac{1 \km}{s_c} \right)^{2.6} \, {\rm
comets} \, ,
\label{comets}
\end{equation}

\ni where $s_c$ is the minimum diameter of a comet. Our order-of-magnitude
estimate compares
favorably with the population of $\sim$$7 \times 10^9$ comets between 30 and 50
AU required to supply the rate of Jupiter-family comets
(\markcite{ld97}Levison \& Duncan 1997).\footnote{The scattered KBO disk has
also been
proposed as an alternative source of short-period comets. \markcite{dl97}Duncan
\& Levison (1997)
estimate that only $6 \times 10^8$ comets are required in the scattered disk to
supply
the observed rate.}

Omitting data from various surveys while preserving the same magnitude coverage
in
the luminosity function
raises the inferred value of $q$ and places most of the mass of the observable
Kuiper Belt
into the smallest objects. \markcite{getal98}Gladman et al. (1998) do not
incorporate
data from JL95, JLC96, LJ98 and CB99.
Their maximum likelihood analysis, which can and does assimilate upper
limit data from \markcite{lj88}Luu \& Jewitt (1988) and \markcite{ld90}Levison
\& Duncan (1990),
concludes that the sky density slope $\alpha = 0.76$. Inserting
this value into equation (\ref{simple}) yields a size index $q = 4.8$.
As a separate example of a shallow slope based on omission of data,
a least-squares fit to the luminosity function
which omits points from JL95, JLC96 and LJ98, and which does not
incorporate upper limit data, yields $q = 4.3$ (see Figure \ref{clftri}b). Both
size indices would place most of the mass of the observable Kuiper Belt into
the smallest objects. Since the size of the smallest object in the distribution
is unconstrained, we cannot estimate the mass
of the Kuiper Belt using these $q$'s. However, for any $q > 4$,
there always exists an $s_{min}$ below which upper
limits for the total cometary Belt mass within 50 AU ($\sim$1.3 $M_{\oplus}$)
are violated.
For values of $q = 4.3$ and $\Sigma (m_R < 27) = 200 \deg^{-2}$ derived from
Figure \ref{clftri}b,
this minimum value for $s_{min}$ is as large as
2 km, and only increases with increasing $q$. Explaining the existence
of such lower cut-off sizes would be problematic.

Our preferred size index, $q = 3.6 \pm 0.1$, is that of a Dohnanyi-like
size distribution for objects having diameters between 50 and 500 km within 50
AU.
However, whether the shape of this distribution results from a catastrophic
collisional
cascade as envisioned in \markcite{d69}Dohnanyi's (1969) scenario is
questionable.
The answer depends on impact strengths, relative velocities, and initial
populations
of KBOs, all of which are poorly constrained. For solid rocky bodies 50--500 km
in diameter,
critical specific energies for shattering and disruption are expected to
increase strongly with size due to self-gravitational compression
\markcite{mr97}(Melosh
\& Ryan 1997, and references therein).
The role of self-gravity is magnified yet further if bodies consist
predominantly
of weaker ices. Whatever their composition, we would not expect
Dohnanyi's (1969) derivation to apply to objects as large as those observed,
since the derivation assumes that impact strengths are independent of size.
Relative velocities required for fragmentation and dispersal of
solid rocky bodies $\sim$100 km in size demand KBO eccentricities
and inclinations exceeding 0.3; the
actual history of the velocity dispersion is unknown. If KBOs consist of solid
rock and
relative velocities are sufficiently high for disruption and dispersal upon
impact,
we estimate that lifetimes against catastrophic dispersal of targets $\sim$100
km in diameter
exceed the age of the Solar System by a factor of $\sim$150 if
projectiles are drawn from the present-day Kuiper Belt.
This estimate agrees with that of \markcite{s95}Stern (1995);
see his Figure 2.
Shaping the population of objects having sizes 50--500 km by
catastrophic collisions
would require a primordial Belt orders of magnitude more populous
than what is observed today.

\subsection{A Kuiper Cliff at 50 AU?}
\label{cliff}

To date not one member of the classical Kuiper Belt has been discovered
beyond 50 AU, despite observational advances in limiting magnitude and
theoretical assurances that the region is dynamically stable.
\markcite{getal98}Gladman et al. (1998)
have addressed this issue and concluded that the present sample of $\sim$100
KBOs is marginally
large enough to expect detection of such bodies. Here we confirm
and elaborate upon their results.

We assume the Kuiper Belt begins at an inner edge $a_{min}$, and that the
number
density of objects (number per volume) decreases with distance $a$ as a
power law with index $\beta$:

\begin{equation}
dN (s, a) \propto a^{-\beta} \, s^{-q} \, ds \, .
\end{equation}

\ni For a surface density (number per disk face area) appropriate to
the minimum-mass outer solar nebula, the index $\beta$ may plausibly
take values of $\sim$2--3, depending on how quickly random
eccentricities $e$ and inclinations $i$ decay with heliocentric distance.
In a field of limiting magnitude $m$,
the sky density of objects (number per projected sky solid angle)
located beyond distance $a_*$ is proportional to

\begin{equation}
\Sigma \, (a > a_*) \propto \int_{s_m(a_*)}^{s_{max}} \int_{a_*}^{a_m(s)}
a^{2-\beta} s^{-q} \, ds \, da \, ,
\end{equation}

\ni where $s_m(a_*)$ is the size of the smallest object which can just be seen
at $a_*$,
and $a_m(s) = a_* \sqrt{s / s_m(a_*)}$ is the maximum distance out to which an
object of
size $s$ can be seen. One immediate consequence of a Belt having distance and
size indices
considered here is that the faint end of the luminosity function is dominated
by small nearby
objects rather than large distant ones. Extending the limiting magnitude of a
visual
survey inherently achieves greater dynamic range in observable sizes than in
distances
because reflected fluxes decrease as $s^2 / a^4$. The greater sensitivity to
size is
compounded by the shapes of the distributions;
$\Sigma \propto s^{1-q} \, a^{3-\beta} \approx s^{-2.6} \, a^{0.5}$.
An outer edge to such a Belt at 50 AU (a ``Kuiper Cliff'')
would not significantly flatten the slope of the luminosity function
at faint magnitudes, a point which we shall justify more formally below.

The fraction of objects located beyond $a_*$ is

\begin{equation}
f \equiv \frac{\Sigma \, (a>a_*)}{\Sigma \, (a>a_{min})} = \left(
\frac{a_{min}}{a_*} \right)^{\gamma} \left\{ 1 + O \left[ \left(
\frac{s_m(a_*)}{s_{max}} \right)^{\gamma /2} \right] \right\} \, ,
\end{equation}

\ni where $\gamma = 2q + \beta - 5$ (cf. \markcite{getal98}G98).
The order-of-magnitude correction term is valid for $\beta \leq 3$
and is small for surveys and distributions considered here.\footnote{For
$\gamma \approx 5$ and $s_{max} = 2000 \km$, the correction
term is less than 0.1 for surveys
having limiting magnitudes $m_R \gtrsim 22$. All surveys used to construct
our luminosity function satisfy the latter requirement.}
The fraction $f$ is thus insensitive to the limiting magnitude
of the field. This insensitivity justifies our assertion that a Kuiper Cliff
would not
break the luminosity function at any particular magnitude.
It also allows us to easily estimate how many detections beyond 50 AU
we might expect. For $a_{min} = 30 \AU$, $a_* = 50 \AU$, $s_{max} = 2000 \km$,
$q = 3.6$, and $\beta = 3$
(constant dispersion in $e$ and $i$), the fraction of objects outside 50 AU is
$f \approx 8\%$.
Decreasing the distance index $\beta$ to 2 increases $f$ to 13\%.
In the present total sample of $\sim$100 KBOs, we might therefore expect
$\sim$10 to
reside beyond 50 AU. Eight of these ten would be located between 50 and 70 AU.

While these rough considerations do not convincingly implicate
a Kuiper Cliff, they do argue more strongly against a sudden rise by factors of
3 or more
in the surface number density between 50 and 70 AU (a nearby ``Kuiper Wall'').
Keeping the size distribution fixed and multiplying the surface density
by 3 beyond 50 AU would demand that $\sim$25\% of all detected classical KBOs
reside in such a wall, in contrast to the 0\% found to date.

Nonetheless, there are a number of ways the present lack of detections may
still accord with a massive outer classical Belt. The size distribution of
objects
may change dramatically past 50 AU. For instance, it might be that only a
few large objects exist between 50 and 70 AU. This may plausibly be the
result of runaway accretion unimpeded by the presence of Neptune.
Alternatively, if only objects smaller than
$\sim$30 km populate the outer Belt, detecting them is a task better suited
to occultation surveys than to searches relying on reflected light.

\section{SUMMARY}
\label{summary}

Our main results are as follows.

\begin{enumerate}

\item We discovered 2 new Kuiper Belt Objects in a single Keck LRIS field.
One object at $m_V = 25.5$ was found by blinking individual frames.
It lies at a heliocentric distance of $R \approx 33 \AU$ and has
a diameter $s = 56 \, (0.04/p_V)^{1/2} \, \km$. Another object
at $m_V = 27.2$ was discovered by blinking shifted and co-added frames.
For this second object, $R \approx 44 \AU$ and $s = 46 \, (0.04/p_V)^{1/2} \,
\km$.

\item We pooled all surveys to construct the cumulative luminosity function
from $m_R = 20$ to 27 (Figure \ref{clfuni}). At the faintest observed
magnitude,
$\Sigma (m_R \lesssim 26.9) = 53 \pm 20 (1\sigma)$ objects/square degree.
The best-fit slope is $\alpha = 0.52 \pm 0.02$, where
$\log_{10} \Sigma (< m_R) = \alpha (m_R - m_0)$. Differences in $\alpha$
reported in the literature are due mainly to which survey data are
incorporated. Values of $\alpha > 0.6$
require the omission of surveys by JL95, JLC96, and LJ98.

\item Our KBO luminosity function is consistent with a power-law
size distribution with differential size index $q = 3.6 \pm 0.1$
for objects having diameters 50--500 km within 50 AU.
The distribution is such that the smallest objects possess most
of the surface area, but the largest bodies contain the bulk of
the mass. By extrapolating outside the observed range of sizes,
we estimate to order-of-magnitude that $0.2 M_{\oplus}$ and
$1 \times 10^{10}$ comet progenitors lie between
30 and 50 AU. Though our estimated size index is that
of a Dohnanyi-like distribution, the interpretation
that catastrophic collisions are responsible
is questionable. Impact strengths against catastrophic disruption
and dispersal probably increase strongly with size for objects
greater than $\sim$10 km in diameter, whereas the derivation
by \markcite{d69}Dohnanyi (1969) assumes impact strength to be
independent of size. Lifetimes against catastrophic dispersal
of KBOs having diameters 50--500 km exceed the
age of the Solar System by at least 2 orders of magnitude
in the present-day Belt, assuming bodies consist of competent rock.

\item A greater than threefold rise in the surface
density of the Kuiper Belt just beyond 50 AU
would imply that more than 25\% of detected objects
lie outside that distance, assuming objects are
distributed similarly in size at all distances.
The absence of detections past 50 AU in the present
sample of $\sim$100 KBOs argues against this picture.
A massive outer Belt may still be possible
if only a few large objects exist between 50 and 70 AU, or if
only objects smaller than $\sim$30 km exist in the outer
Belt.
\end{enumerate}

\acknowledgements
Data were obtained at the W. M. Keck
Observatory, which is operated as a scientific partnership
among the California Institute of Technology, the Universities of California,
and the National Aeronautics and Space Administration. The observatory was
made possible by the generous financial support of the W. M. Keck Foundation.
We thank Jane Luu for providing detection efficiencies for the LJ98 data,
Peter Goldreich and Sarah Stewart for helpful discussions, and
an anonymous referee for a careful reading of this manuscript.
E.C. gratefully acknowledges support from an NSF Graduate Research Fellowship.

\end{document}